# Exploiting Ligand Additivity for Transferable Machine Learning of Multireference Character Across Known Transition Metal Complex Ligands

Chenru Duan[1,2], Adriana J. Ladera[1], Julian C.-L. Liu[1], Michael G. Taylor[1], Isuru R. Ariyarathna[1], and Heather J. Kulik[1,*]

[1]*Department of Chemical Engineering, Massachusetts Institute of Technology, Cambridge, MA 02139*

[2]*Department of Chemistry, Massachusetts Institute of Technology, Cambridge, MA 02139*

ABSTRACT: Accurate virtual high-throughput screening (VHTS) of transition metal complexes (TMCs) remains challenging due to the possibility of high multi-reference (MR) character that complicates property evaluation. We compute MR diagnostics for over 5,000 ligands present in previously synthesized transition metal complexes in the Cambridge Structural Database (CSD). To accomplish this task, we introduce an iterative approach for consistent ligand charge assignment for ligands in the CSD. Across this set, we observe that MR character correlates linearly with the inverse value of the averaged bond order over all bonds in the molecule. We then demonstrate that ligand additivity of MR character holds in TMCs, which suggests that the TMC MR character can be inferred from the sum of the MR character of the ligands. Encouraged by this observation, we leverage ligand additivity and develop a ligand-derived machine learning representation to train neural networks to predict the MR character of TMCs from properties of the constituent ligands. This approach yields models with excellent performance and superior transferability to unseen ligand chemistry and compositions.

# 1. Introduction

Virtual high-throughput screening (VHTS)[1-8] and machine learning (ML)-accelerated chemical discovery[9-17] with approximate density functional theory (DFT) have started to address the combinatorial challenges[18-23] in discovering and designing functional molecules and materials. Despite the balanced trade-off in computational cost and accuracy, DFT can fail prominently[24-30] for promising materials, such as the chemical space of transition metal complexes (TMCs). TMCs may have strong multi-reference (MR) character due to near-degenerate orbitals,[31,32] which cannot be accurately described in DFT due to the single-reference (SR) nature of the non-interacting wavefunction.[33] In addition, most density functional approximations (DFAs) are designed and benchmarked with a focus on main group systems.[34-36] As a result, "Jacob's ladder"[37], which suggests that DFAs lying at higher rungs will be more accurate, holds well for organic molecules but often fails to describe performance on TMCs.[4,24,26,33,38] To maintain data fidelity in transition metal chemical discovery, it is vital to determine whether a system contains strong MR character and choose an appropriate method in VHTS.[39,40]

Many studies have devised MR diagnostics[31,41-52] based on distinct properties (e.g., occupations or atomization energies) and levels of theory to quantify the degree of MR character. Of these MR diagnostics, those derived from wavefunction theory (WFT) are generally more predictive than those derived from DFT.[53] However, WFT calculations scale at least $N^4$ with the system size and thus are impractical in VHTS of transition metal chemistry. Moreover, certain MR diagnostics have been observed to be more transferable over different chemical spaces (i.e. organic molecules vs. TMCs) than others.[54-56] Therefore, data-driven approaches[57-61] have been developed to bridge the gap between DFT- and WFT-based diagnostics[53], making system-specific MR/SR decisions for method selection[39], and automating the active space selection for MR WFT calculations[57,60,62].

Strong MR character in TMCs still poses challenges for VHTS. Most studies that investigate the MR character of TMCs focus on model systems with a few ligands. This is likely due to the prohibitive computational cost of studying a large data set of realistic, synthetically accessible TMCs with multiple WFT calculations. In addition, despite the fact that the MR character of both TMCs and organic molecules have been studied in the literature, the relationship between the MR character of a TMC and that of its constituent ligands remains unknown.

In this work, we study TMCs and thousands of their constituent ligands that are known to be synthetically accessible by virtue of their presence in the Cambridge Structural Database (CSD)[63]. We devise a strategy to quantify the total MR character of TMCs based on ligand-derived measures of MR character. This approach makes tractable the diagnosis of high MR character in TMCs. To accomplish this task over a large set of known ligands, we first introduce an iterative approach for ligand charge assignment on CSD ligands based on all charge and oxidation state information available within the CSD. We then perform high-throughput calculations to obtain MR diagnostics for 5,163 synthetically accessible ligands. For these ligands, we observe that ligand MR character linearly correlates with the inverse value of the averaged bond order over all bonds in the ligand. We demonstrate that the concept of ligand additivity holds for MR character, which suggests that the MR character of a TMC can be inferred from the sum of the MR character of the ligands. Leveraging this ligand additivity, we develop a new ligand-derived ML representation to train ANNs to predict TMC MR character.

These models demonstrate excellent performance and superior transferability to unseen ligand chemistry and compositions in comparison to alternative approaches.

## 2. Data Sets

We searched the CSD[63] version 5.41 (Nov. 2019) + 3 data updates to find octahedral mononuclear complexes. This query returned 28,006 complexes that have user-defined charge and oxidation states. We used molSimplify[64,65] to deconstruct these complexes into metals and their corresponding ligands, which resulted in 9,279 unique ligands. Duplicate ligands were eliminated using our previous molecular graph determinant[66] approach. We applied our iterative CSD-consistent charge assignment approach on these complexes and successfully assigned the charge of 8,805 ligands (see *Results and Discussions 4.1*). We excluded ligands that have more than 25 heavy atoms and those that contain uncommon chemical elements, as judged by their low abundance (< 0.1%) in the ligands of mononuclear octahedral complexes in CSD. After removing these ligands from the overall set, we were left with 6,131 ligands in total (Supporting Information Table S1).

We compared our CSD-consistent charge with charge assignment using the octet rule[67]. Only 5,905 ligands can be assigned a charge using the octet rule, because the remaining examples lack bond order information. We identified 5,713 cases where our assigned charge matches an octet rule charge and 192 cases where the two charges differed. We manually inspected all 192 cases where the charges differed and observed that the most common reason for a mismatch was due to incorrect treatment of radicals or ligands containing multiple atoms that have variable oxidation states (e.g., B and P) in the octet rule assignment (Supporting Information Figure S1). We confidently identified 57 of these 192 cases as radicals or diradicals by inspecting the corresponding literature. A complete list of these ligands is provided in a .csv file in the Supporting Information. We then computed the MR diagnostics (see *MR diagnostics calculations*) of the 5,713 closed-shell ligands and 57 radicals or diradicals following our established procedure[53]. If we could not successfully compute all 14 MR diagnostics (e.g., due to self-consistent field convergence failure), we removed that ligand from the data set (Supporting Information Table S2). Our final data set contains 5,163 ligands from octahedral mononuclear CSD complexes for which we have ascertained charges and computed MR diagnostics, which we refer to as *OctLig*.

To investigate the effect of mixing ligands with distinct MR character, we also selected ligands from the *OctLig* set that have the strongest or weakest MR character for their respective denticities of 1, 2, 5, or 6, which corresponds to a set of 8 ligands in total (Supporting Information Figure S2). We combined these ligands with Fe centers to make octahedral complexes. We considered both Fe(II) and Fe(III) in high-spin (HS) (i.e., quintet or sextet) or low-spin (LS) (i.e., singlet or doublet) states and all possible ligand symmetries, which produces a set of 172 transition metal complexes that we refer to as *extTMC* (Supporting Information Table S3).

We also assembled 1,000 octahedral complexes from monodentate ligands in the *OctLig* set under the constraints that the complex consists of up to 2 unique ligands, that it was no more than 40 atoms in size, and had a net charge of -4 to +2. Each complex contains a mid-row transition metal (Cr, Mn, Fe, or Co) in oxidation state II or III and in its HS or LS state (Supporting Information Table S4). We performed DFT geometry optimizations on these 1,000 TMCs, and 855 of them converged to suitable octahedral geometries (Supporting Information Table S5). We refer to this set of 855 complexes from diverse CSD ligands as *divTMC*.

**3. Methods**

### 3.1 DFT Geometry Optimizations.

Density functional theory (DFT) geometry optimizations with the B3LYP[68-70] global hybrid functional were carried out using a developer version of the graphical processing unit (GPU)-accelerated electronic structure code TeraChem[71-73]. The LANL2DZ effective core potential[74] basis set was used for metals, Br, and I, and the 6-31G* basis for all other atoms. Singlet spin states were calculated with the spin-restricted formalism while all other calculations were carried out in a spin-unrestricted formalism. All DFT calculations employed level shifting[75] of 0.25 Ha to both majority- and minority-spin virtual orbitals to aid self-consistent field (SCF) convergence. All geometry optimizations were performed with the L-BFGS algorithm in translation rotation internal coordinates (TRIC)[76] to the default tolerances of $4.5 \times 10^{-4}$ Hartree/bohr for the maximum gradient and $1 \times 10^{-6}$ Hartree for the energy change between steps.

For the *OctLig* set, we selected the initial ligand geometry as the structure that produced the lowest B3LYP single-point energy among the multiple copies of the ligand present in the CSD. In the subsequent geometry optimization of that structure, we only optimized the location of H atoms, which are not necessarily resolved in the crystal structure, while we froze all heavy atoms. For the *extTMC* and *divTMC* sets, we assembled transition metal complex initial geometries using molSimplify[64,65] to combine the optimized ligand structures from the *OctLig* set with transition metals in octahedral geometries. Geometry checks[77,78] were applied to eliminate optimized TMC structures that deviated from the expected octahedral shape following previously established metrics without modification[77,78] (Supporting Information Table S5).

### 3.2 MR Diagnostic Calculations.

Following our prior studies[39,79], we calculated 14 MR diagnostics[31,41-50] for the *CSDLig* set using ORCA 4.2.1 with the cc-pVDZ basis set on the H, B, C, N, O, and F and the cc-pVTZ basis set on all other atoms (Supporting Information Tables S6–S7). These DFT-based and WFT-based diagnostics are all explicitly calculated, unlike some previous work where we trained ML models to predict the higher cost WFT-based diagnostics from a combination of descriptors and DFT-based diagnostics.[79] We used our established workflow[39,54,79] to ensure convergence to a consistent electronic state when evaluating different MR diagnostics (Supporting Information Text S1 and Figure S3). We used the restricted formalism for closed-shell species and restricted open-shell formalism otherwise. We chose the percentage of correlation energy recovered by CCSD compared to CCSD(T) (i.e., %$E_{corr}$[(T)]) as a figure of merit, since we observed good correspondence of %$E_{corr}$[(T)] and %$E_{corr}$[T] (i.e., the percentage of correlation energy recovered by CCSD compared to CCSDT) in both organic molecules[53] and TMCs[54] from our previous work. For the *extTMC* and *divTMC* sets, we used DLPNO-CCSD(T) as a proxy for the canonical CCSD(T) to compute %$E_{corr}$[(T)] due to the sufficient accuracy of DLPNO-CCSD(T) and high computational cost of canonical CCSD(T)[80] on larger molecules (Supporting Information Table S8).

**3.3 ML Model Training.**

We first partitioned the *divTMC* set with a random 80%/20% train/test split, with 20% of the training data (i.e., 16% overall) used as the validation set for hyperparameter selection using HyperOpt[81] with 200 evaluations (Supporting Information Table S9). To examine the transferability of our newly developed ligand diagnostic revised autocorrelation functions (RACs), we further applied two non-random partition schemes on the *divTMC* set based on ligand identity to ensure distinct unseen chemistry in the test set. First, 50 ligands (out of 344 ligands in total) were chosen, and all complexes that contain any of those 50 ligands were partitioned into a test set (Supporting Information Table S10). Second, all complexes that contained any ligand with a 3p (i.e., period 3) connecting atom (e.g., Si, S, P, or Cl) were partitioned into the test set (Supporting Information Table S10). For these two partitioning schemes, we adopted the hyperparameters from hyperparameter optimization for the random partition (Supporting Information Table S9). All ANN models were trained up to 2,000 epochs with the Adam optimizer using Keras[82] with Tensorflow[83] as a backend, and dropout, batch normalization, and early stopping to avoid over-fitting.

## 4. Results and Discussion

### 4.1 Consistent Ligand Charge Assignment Using Collective Information in the CSD.

There are 28,006 octahedral mononuclear transition metal complexes deposited in the Cambridge Structural Database (CSD)[63] (version 5.41 (Nov. 2019) + 3 data updates) that have user-labeled complex charges and metal oxidation states. This information facilitates computational studies of their properties (e.g., HOMO–LUMO gap) that can be obtained from DFT.[84] Similar by-ligand charge information, however, is absent for the ligands that comprise these CSD complexes, preventing large-scale computational screening of individual synthetically accessible ligands. Thus, to assess a ligand's MR character and how it correlates to the MR character of TMCs that contain it, we need a method that assigns charges of individual ligands. We thus developed an iterative approach to obtain the ligand charges that is consistent over multiple copies of the ligand in the CSD (Supporting Information Text S2). This strategy applies the constraint that the total complex charge should be the sum of the metal oxidation state and the sum of all constituent ligand charges (Supporting Information Text S2). We start with homoleptic complexes (i.e., with only one unique ligand) and directly derive the charge of the ligand involved. Next, we search for complexes that only have one ligand with unknown charge while others have their charges assigned from a previous iteration of our algorithm. This process continues until we cannot obtain charges for any new ligands (Figure 1). This iterative approach has several advantages. First, the deduced ligand charges are consistent across the CSD and are thus likely to be correct. Second, the number of ligands with known charge will increase as more structures are deposited in the CSD (Supporting Information Figure S4). Lastly, we can use the same logic to obtain either the metal oxidation state or the total complex charge for complexes when this information is not available in the deposited CSD structure.

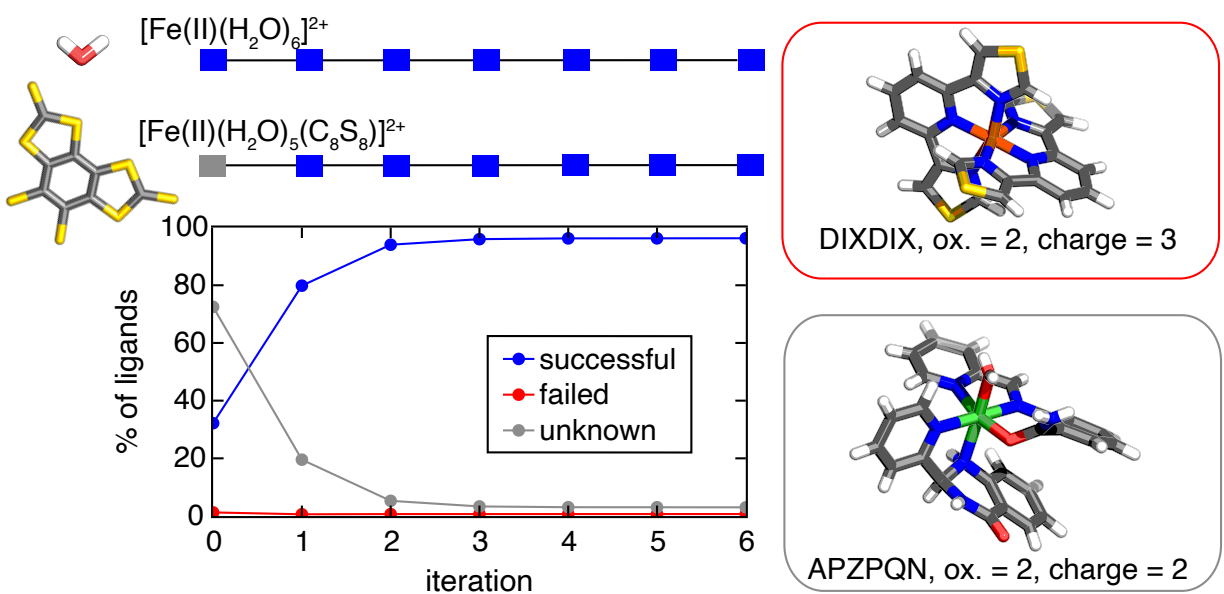

**Figure 1.** (left) The percentage of ligands versus the iteration number in the CSD-consistent charge assignment process (bottom left). Ligands are categorized into three groups: successfully assigned (blue), failed assignment due to conflicting information in CSD entries (red), and unknown (gray). An illustration of the iterative CSD-consistent charge assignment for the example of two complexes, $[Fe(II)(H_2O)_6]^{2+}$ and neutral $Fe(II)(C_8S_8)(H_2O)_4$, is also shown (top). (right) Examples where CSD-consistent charge assignment will fail: a homoleptic Fe(II) complex (refcode: DIXDIX) with two identical tridentate ligands that leads to a non-integer charge assignment for each tridentate ligand (i.e., 0.5, top) and a Ni(II) complex (refcode: APZPQN) that is comprised of two ligands, both of which rare in the CSD. Atoms are colored as follows: orange for Fe, green for Ni, yellow for S, gray for C, blue for N, red for O, and white for H.

With this iterative approach, we successfully obtain CSD-consistent charges for 96.3% of ligands seen in octahedrally coordinated complexes in the CSD (Figure 1). The remaining 3.7% of ligand failures are mostly due to mislabeling or lack of information in the CSD. For 0.7% of the ligands, we failed to assign the charge due to potential mislabeling of the total complex charge and/or the metal oxidation state by the user who deposited the crystal structure. For example, an Fe(II) complex (refcode: DIXDIX) is reported with a total complex charge +3 but contains two identical tridentate ligands, leading to a fractional ligand charge assignment (i.e., 0.5) for each tridentate ligand (Figure 1). The remaining 3.0% of the ligands are rarely present in the CSD and thus their charges are challenging to determine through inference. For example, a Ni(II) complex (refcode: APZPQN) contains two ligands that only appear once in all octahedral complexes in the CSD, leading difficulties for their CSD-consistent charge assignment (Figure 1). For the ligands where we are able to assign a CSD-consistent charge, we obtain good agreement (96.7%) with charges derived from the octet rule.[67] When the CSD-consistent charge disagrees with the octet rule charge, the ligands are either radicals, diradicals, or contain multiple atoms that have variable oxidation states (e.g., B and P), where the octet rule is expected to fail (Supporting Information Figure S1).

**4.2 Statistics of MR Character for Over Five Thousand Ligands in the CSD.**

We next analyzed the MR character, as judged by $\%E_{\text{corr}}[(T)]$, for the 5,163 synthetically accessible ligands in the *OctLig* set (see *Methods* and *Data Sets*). We find a strong linear correlation (Pearson's $r$=0.79) between the inverse value of the averaged bond order (BO) over all bonds for each ligand and its corresponding $\%E_{\text{corr}}[(T)]$ (Figure 2). This correlation holds regardless of the size of the ligand in the *OctLig* set (Supporting Information Figure S5). Small ligands with triple bonds and thus a very small inverse average bond order, such as CS, CN⁻, and $C_4H^-$, have the strongest MR character (Supporting Information Table S11). Trends in %$E_{corr}$[(T)] over this set are in accordance with our prior work, where we studied stretched organic molecules[39,53] and model transition metal complexes[54]. Here, we again observe poor linear correlations between %$E_{corr}$[(T)] and the 14 computed MR diagnostics for this set of synthetically accessible ligands (Supporting Information Figures S6–S7). This observation suggests that the lack of consensus between different MR diagnostics is general across a range of chemistries and that simple heuristics such as bond order can be more predictive of MR figures of merit such as %$E_{corr}$[(T)] than computationally costly WFT-derived diagnostics.

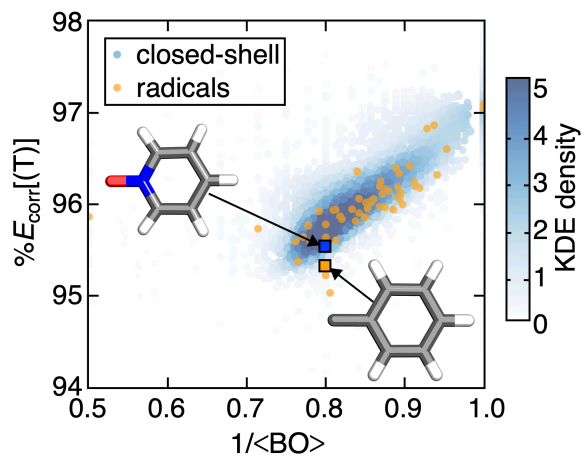

**Figure 2.** %$E_{corr}$[(T)] versus the inverse average bond order for CSD ligands in the *OctLig* set. The closed-shell ligands are colored by the kernel density estimation (KDE) density values, as indicated by the color bar, while the radicals are colored in orange. Two example ligands both with an inverse average bond order (BO) of 0.8 and the same size are shown: phenylmethylidyne ($C_7H_5$), a diradical and closed-shell pyridine-N-oxide ($C_5H_5NO$). Atoms are colored as follows: gray for C, blue for N, red for O, and white for H.

Due to the good linear correlation between %$E_{corr}$[(T)] and the inverse average BO, ligands with higher denticities have weaker MR character, because they tend to be more saturated than ligands with lower denticities (Figure 3). This observation suggests that single-site catalysts with high-denticity ligands are preferable targets for VHTS not only due to their ease of their synthesis and introduction of steric effects that can be beneficial "knobs" to tune in catalyst design[85,86] but also due to their simpler electronic structure. Over the 5,163 unique ligands from the *OctLig* set, our algorithm identifies 57 radical ligands, which we confirm by literature analysis (details are provided in a .csv file in the Supporting Information .zip file). Surprisingly, these radical ligands do not have increased MR character relative to the closed-shell singlets when their inverse average bond orders are comparable (Figure 2). For example, phenylmethylidyne ($C_7H_5$), a diradical ligand with an inverse average bond order 0.8, has similar %$E_{corr}$[(T)] compared to the closed-shell pyridine-N-oxide ($C_5H_5NO$), which has the same inverse average bond order, size, and similar molecular weight (Figure 2 and Supporting Information Figure S8).

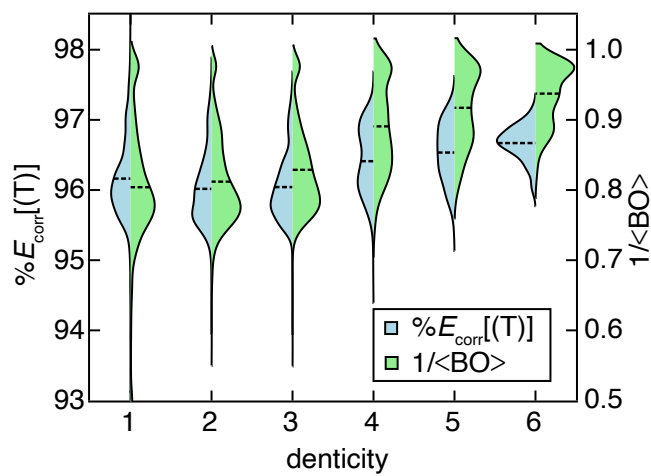

**Figure 3.** Split violin plot of %$E_{corr}$[(T)] (blue) and the inverse average bond order (green) for CSD ligands in the *OctLig* set at different denticities. In each split violin, the mean value is shown as a dashed line.

**4.3 Incorporating Additivity of MR character for TMCs Improves ML Transferability.**

We next investigated the relationship between the MR character of individual ligands and the MR character of the transition metal complexes (TMCs) that they comprise. We started with a controlled data set of Fe(II) complexes, *extTMCs*, in which all TMCs are constructed from ligands that have either strong (e.g., CS) or weak (e.g., $H_2O$) MR character (see *Data Sets*). By constructing TMCs in this manner, we investigate how TMC MR character changes if we exchange a ligand with strong MR character for a ligand that has weak MR character. As we increase the number, n, of CS ligands in octahedral LS Fe(II)(CS)$_n$(H$_2$O)$_{6-n}$, we observe a monotonically decreasing and near-linear trend for the DLPNO-CCSD %$E_{corr}$[(T)] (Figure 4). This indicates that a complex made up of ligands with strong MR character itself has high MR character. Different isomers of the complex with the same chemical composition (i.e., n) have comparable DLPNO-CCSD %$E_{corr}$[(T)], and differences among isomeric TMCs are smaller than between TMCs comprised of different ligands (Figure 4). This ligand additivity effect has been previously explored for the spin-splitting energy of TMCs[87] but has not yet been observed in other properties of TMCs. Ligand additivity holds for nearly all MR diagnostics considered in this work and for other series of TMCs (e.g., HS Fe(II)(CS)$_n$(H$_2$O)$_{6-n}$) in the *extTMCs* set (Supporting Information Figures S9–S12). Notably, the only exception to the observed increasing MR character is the $r_{ND}$ diagnostic. With this diagnostic, ligand additivity does not appear to hold or, in some cases, follows the opposite of the expected trend (Supporting Information Figures S9–S10). This unexpected trend cannot be attributed to size-dependence of the diagnostic because $r_{ND}$ is designed to be size intensive[88], and the trend with increasing number of strongly MR ligands is in opposing directions for the LS and HS cases of Fe(II)(CS)$_n$(H$_2$O)$_{6-n}$ (Supporting Information Figures S9–S10). That is, sometimes $r_{ND}$ increases with increasing MR character of individual ligands, sometimes it decreases, and other times it lacks monotonic behavior. Since all other diagnostics and the figure of merit (i.e., DLPNO-CCSD %$E_{corr}$[(T)]) do follow the expected trend of increasing overall MR character with more ligands that have individually higher MR character, we conclude that the nature of $r_{ND}$ (i.e., the ratio of non-dynamical to total correlation) intrinsically fails to capture additivity effects in MR character. The $r_{ND}$ diagnostic also sometimes shows larger sensitivities to isomer choice than to the character of the ligands, which diverges from the other computed diagnostics and figure of merit (Supporting Information Figure S10).

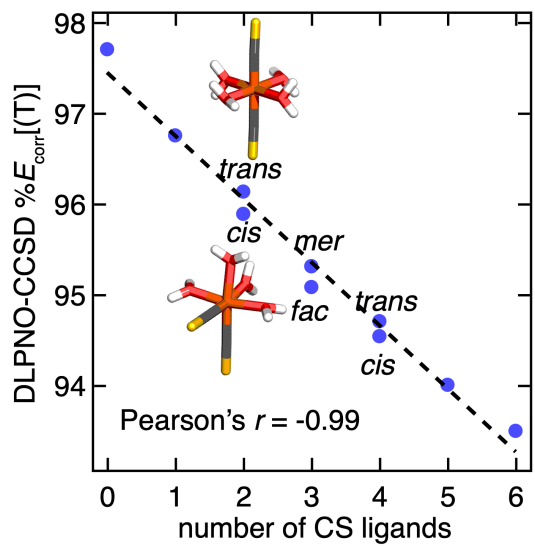

**Figure 4.** $\%E_{corr}[(T)]$ computed with DLPNO-CCSD(T) versus the number of CS ligands in LS Fe(II)(CS)$_n$(H$_2$O)$_{6-n}$. The type of isomer is annotated for n=2, 3, and 4 with example *trans* and *cis* structures shown in the inset. A best-fit dashed line is shown for the 10 complexes. Atoms are colored as follows: orange for Fe, gray for C, yellow for S, red for O, and white for H.

With our previously developed extended RACs (eRACs)[89,90], we next built ANN models to directly predict the DLPNO-CCSD $\%E_{corr}[(T)]$ figure of merit because individual diagnostics do not necessarily correlate well with this quantity. Such models are envisioned to be useful in VHTS to identify the "safe islands"[91] where DFT is a suitable level of theory or, alternately, identify where WFT calculations are necessary. The eRAC representation consists of sums of products and differences of atomic properties on a 2D molecular graph of a TMC for pairs of atoms separated by a certain bond depth (Supporting Information Text S3). The properties include the electronegativity ($\chi$), nuclear charge ($Z$), topology ($T$), covalent radius ($S$), and identity ($I$) from the original RACs representation[90] that are extended in eRACs to also include group number ($G$). For example, the feature among eRACs that has the best linear correlation with DLPNO-CCSD %$E_{corr}$[(T)] (Pearson's $r$=0.59), $_{eq}^{lc}G'_1$, is the average of the difference in group number between the metal-coordinating atom of the equatorial ligands and their nearest neighbor non-metal atoms (Figure 5). We observe good performance for the ANN model using eRACs as inputs on the set-aside test data ($R^2$=0.95) for a random train/test partition of the *divTMC* set made of diverse combinations of metals and monodentate ligands. This performance is similar to our previous observations when applying RACs and eRACs for predicting properties of TMCs, such as spin-splitting energy[89], metal–ligand bond length[66], and MR diagnostics[91] (Figure 6). However, we find that this eRACs-based ANN model is not sufficiently transferable. The eRACs/ANN model performance is worsened if tested on ligands that were held out in a group split ($R^2$=0.81) or chemical elements (i.e., unseen 3p elements when only 2p elements were in the training set, $R^2$=0.51) that were not present in the training set (Figure 6 and see *Data Sets*). We thus next investigated whether additivity could be exploited to devise new descriptors from a ligand-focused perspective that would increase model transferability to unseen chemistry.

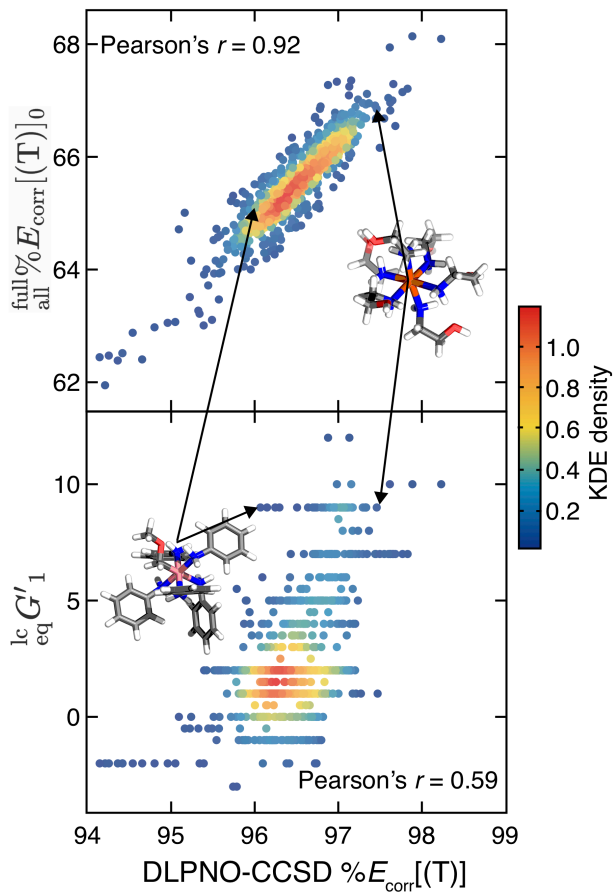

**Figure 5.** $^{full}_{all}\%E_{corr}[(T)]_0$ (top) and $^{lc}_{eq}G'_1$ (bottom) versus $\%E_{corr}[(T)]$ computed with DLPNO-CCSD(T) for the *divTMC* set. Complexes are colored by the kernel density estimation (KDE) density values, as indicated by the color bar. Example complexes, LS Co(III)(C$_6$H$_8$N)$_5$(C$_2$H$_5$O) and HS Fe(II)(C$_2$H$_7$NO)$_6$, are shown in insets. Atoms are colored as follows: pink for Co, orange for Fe, gray for C, blue for N, red for O, and white for H.

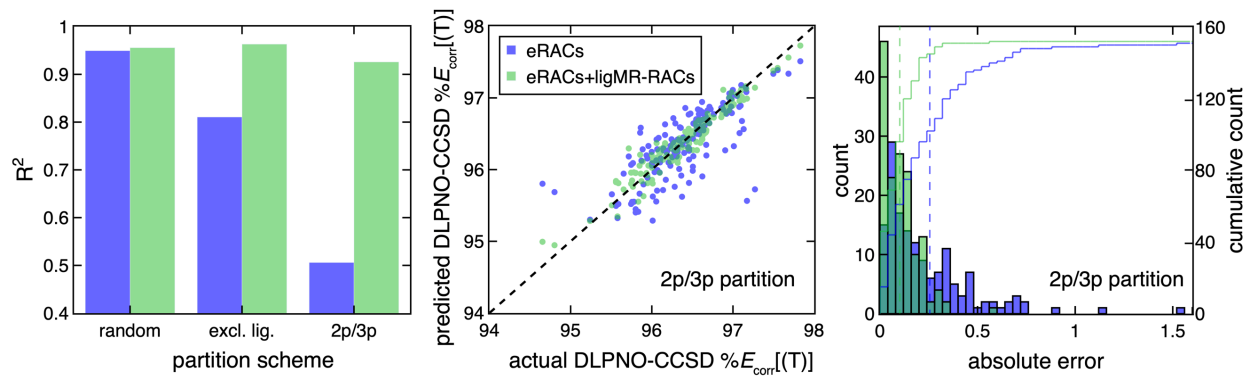

**Figure 6.** (left) R$^2$ for ML-predicted DLPNO-CCSD $\%E_{corr}[(T)]$ with three partitioning schemes (from left to right): a random split of the data, a grouped split that excludes ligands (excl. lig.) from the training set, and removal of all 3p-containing ligands from the training set and placing them in the test set. (middle) Predicted versus actual DLPNO-CCSD $\%E_{corr}[(T)]$ on the set-aside

test data points for the 2p/3p partition scheme. A black dashed parity line is also shown. (right) Distributions of absolute test errors for DLPNO-CCSD %$E_{\text{corr}}$[(T)] (%, bins of 0.04) with the MAE annotated as vertical bars and the cumulative count shown in green or blue according to the axis on the right. In all cases, the results of using only eRACs (blue) and both eRACs and ligMR-RACs (green) as model inputs are shown.

To improve model transferability, we designed a new set of descriptors to exploit the ligand additivity effects in MR character in TMCs, which we refer to as ligMR-RACs. Specifically, we adapt RACs to encode the MR character of a TMC through the MR diagnostics of its fragments (i.e., metal center and ligands). Because most of the MR diagnostics cannot be localized to individual atoms, we first convert the 2D molecular graph to a fragment-based graph, where the MR diagnostics of each fragment (i.e., ligand) have been pre-computed (Supporting Information Figure S13). To compute ligMR-RACs for a TMC, we follow the same procedure as in RACs, but we instead operate on the ligand-based graph and replace the atomwise properties in RACs with these ligand MR diagnostics. Because the maximum bond depth ($d$) is 2 in the fragment-based graph for all TMCs, we truncate ligMR-RACs at depth 2 as well, resulting in 101 ligMR-RACs in total (Supporting Information Text S3). For example, $^{\text{full}}_{\text{all}}\%E_{\text{corr}}[(T)]_0$ is the sum of the squares of %$E_{\text{corr}}$[(T)] for the metal and ligands in a TMC. This quantity should correlate well to DLPNO-CCSD %$E_{\text{corr}}$[(T)] of the whole complex if ligand additivity holds perfectly. Indeed, we observe an extremely good linear correlation between $^{\text{full}}_{\text{all}}\%E_{\text{corr}}[(T)]_0$ and the DLPNO-CCSD %$E_{\text{corr}}$[(T)] for TMCs (Pearson's $r$ = 0.92) in the *divTMC* set (Figure 5 and see *Data Sets*). This greater linear correlation of the ligMR-RAC $^{\text{full}}_{\text{all}}\%E_{\text{corr}}[(T)]_0$ compared to $^{\text{lc}}_{\text{eq}}G'_1$ can also be understood by noting that the former does a better job of distinguishing between TMCs that have different degrees of MR character. For example, LS Co(III)(C$_6$H$_8$N)$_5$(C$_2$H$_5$O) and HS Fe(II)(C$_2$H$_7$NO)$_6$ have the same $^{\text{lc}}_{\text{eq}}G'_1$ although their DLPNO-CCSD %$E_{corr}$[(T)] values differ (i.e., 96.1 vs 97.5). Meanwhile, this large difference in DLPNO-CCSD %$E_{corr}$[(T)] of these two complexes can be distinguished using the $^{full}_{all}$%$E_{corr}$[(T)]$_0$ descriptor (Figure 5).

Encouraged by the good linear correlations between ligMR-RACs and the DLPNO-CCSD %$E_{corr}$[(T)] of TMCs and the predictive power of eRACs on TMC properties, we built ANNs to predict DLPNO-CCSD %$E_{corr}$[(T)] with a combined feature set of eRACs and ligMR-RACs. Notably, ligMR-RACs can be precomputed on a pool of ligands at low cost and the number of ligands is smaller than the TMC design space size by orders of magnitude.[40] Thus, ligMR-RACs add only minor computational overhead in ML-accelerated chemical discovery. On a random train/test partition of the *divTMC* set, we only observe a marginal improvement (i.e., 20% reduction in MAE and increase to $R^2$=0.96 for the set-aside test set) for DLPNO-CCSD %$E_{corr}$[(T)] predictions after complementing eRACs with ligMR-RACs, likely due to the fact that eRACs already exhibit good performance ($R^2$=0.95, Figure 6, Supporting Information Figure S14). We next considered whether this combined feature set improves transferability where the eRACs/ANN model performance had been poor.

To test our model transferability, we returned to the two non-random partition schemes on the *divTMC* set based on ligand identity to ensure distinct unseen chemistry in the test set (see *Methods*). For a grouped split where specific ligands were excluded from the training set and partitioned into the test set, adding ligMR-RACs significantly improves model performance in comparison to using only eRACs ($R^2$=0.81 to $R^2$=0.96, Figure 6, Supporting Information Figure S14). Notably, adding ligMR-RACs greatly improves model accuracy when the model is only trained on the TMCs containing 2*p* elements but tested on those containing unseen 3*p* elements ($R^2$=0.51 to $R^2$=0.93, Figure 6, Supporting Information Table S12). For models with both eRACs and ligMR-RACs as inputs, their performance is thus robust against each of these data partitioning schemes ($R^2$ ranges from 0.93 to 0.96). In addition, the fraction of training data in the three partition schemes are comparable (i.e., 73%–82%), suggesting that the inclusion of ligMR-RACs to our models allows us to learn DLPNO-CCSD %$E_{corr}$[(T)] comparably with either *in-distribution* (i.e., random partition) or *out-of-distribution* (i.e., excluding ligands and 2$p$/3$p$ partition) test data (Supporting Information Table S10). This observation highlights the transferability facilitated by ligMR-RACs in predicting DLPNO-CCSD %$E_{corr}$[(T)] for TMCs in diverse chemical spaces. As a comparison, we also trained ANN models using only ligMR-RACs. Despite the slightly worse performance in comparison to the models trained on the combined feature set, the ligMR-RAC-only models maintain good transferability we identified with ligMR-RACs. Specifically, these models have rather uniform performance in all three partitioning schemes ($R^2$ ranges from 0.89 to 0.93, Supporting Information Figure S14 and Table S12). This comparison emphasizes the synergistic benefit of using both eRACs and ligMR-RACs in chemical exploration.

## 5. Conclusions

The accurate screening of transition metal complexes is challenged by the fact that they may suffer from varying degrees of strong correlation. Given the large size of transition metal complexes and their constituent ligands, it is nearly impossible to exhaustively screen these complexes with multi-reference wavefunction theory methods, motivating low-cost detection of multi-reference character for large-scale screening. To address this challenge, we devised a strategy to quantify ligand-derived measures of multi-reference character, which is much more tractable. To make this screening possible, we first introduced an iterative approach for ligand charge assignment for an individual CSD ligand based on all charge and oxidation state information available within the CSD. This approach has high degree of success (ca. 96%) for correct charge assignment that is otherwise challenging to deduce with the octet rule. This method is also systematically improvable as the CSD grows in number of complexes. Next, once charges were assigned, we performed high-throughput calculations to obtain MR diagnostics for 5,163 synthetically accessible ligands. Over this set, we found that ligand MR character (i.e., %$E_{\text{corr}}$[(T)]) is highly correlated with the inverse value of the averaged bond order over all bonds in the ligand. Surprisingly, radical ligands do not have increased MR character relative to closed-shell ligands when they have comparable inverse average bond order. We demonstrated that ligand additivity holds for MR character, suggesting that the MR character of a TMC can be inferred from the sum of the MR character of the ligands. Encouraged by this observation, we proposed ligMR-RACs to complement our previously developed eRAC descriptors for predicting %$E_{\text{corr}}$[(T)] for TMCs. Our ANN models trained on this combined set of eRACs and ligMR-RACs demonstrate excellent performance and transferability to unseen ligand chemistry and compositions. While this work has focused on equilibrium structures, further adaptations, such as those we have already demonstrated for ligand-like molecules[39,79], could make it amenable to cases with stretched bonds. The framework itself requires no modification for the distortion of intraligand bonds since the ligMR-RACs directly capture this effect, but additional descriptors to capture metal-ligand bond stretching would likely be needed. We anticipate that our observations of ligand additivity and our new ligMR-RACs representation will be broadly useful for identifying "DFT-safe" regions of transition metal chemical space in virtual high-throughput screening and for motivating adaptation to higher-level methods when strong correlation is detected.

ASSOCIATED CONTENT

**Supporting Information Statement**. Element filtering and removed CSD ligands; Example ligands mislabeled by the octet rule; Ligands and ligand symmetries considered in *extTMC* set; Oxidation and spin state, geometry metrics cutoffs, and MR diagnostics considered in this work; Workflow, calculation parameters of computing MR diagnostics; Range of hyperparameters for ANN optimizations; Pseudocode for iterative CSD-consistent charge assignment; $\%E_{corr}[(T)]$ vs. inverse average bond order; Unsigned Pearson's r matrix and Spearman's r matrix of MR diagnostics and $\%E_{corr}[(T)]$; Structure for example ligands; Ligand additivity for TMCs in *extTMC* set; 2D molecular graph and extended description for RACs and ligMR-RACs; ML model performance. (PDF)

ML model h5 files; optimized CSD ligands and TMC geometries xyz files; data csv files. (ZIP)

AUTHOR INFORMATION

**Corresponding Author**

*email:hjkulik@mit.edu

**Notes**

The authors declare no competing financial interest.

ACKNOWLEDGMENT

The authors acknowledge support by the Department of Energy under grant number DE-SC0018096 as well as a MolSSI fellowship (grant no. OAC-1547580) to C.D.. A.J.L. was supported by the MIT Summer Research Program. Additional support for data set generation was provided by the Office of Naval Research under grant numbers N00014-18-1-2434 and N00014-20-1-2150. This work made use of Department of Defense HPCMP computing resources. This work was also carried out in part using computational resources from the Extreme Science and Engineering Discovery Environment (XSEDE), which is supported by National Science Foundation grant number ACI-1548562. H.J.K. holds a Career Award at the Scientific Interface from the Burroughs Wellcome Fund, an AAAS Marion Milligan Mason Award, and an Alfred P. Sloan Fellowship in Chemistry, which supported this work. The authors thank Aditya Nandy and Adam H. Steeves for providing a critical reading of the manuscript.References

# Supporting Information for

*Exploiting Ligand Additivity for Transferable Machine Learning of Multireference Character Across Known Transition Metal Complex Ligands*

Chenru Duan[1,2], Adriana J. Ladera[1], Julian C.-L. Liu[1], Michael G. Taylor[1], Isuru R. Ariyarathna[1], and Heather J. Kulik[1,*]

[1]Department of Chemical Engineering, Massachusetts Institute of Technology, Cambridge, MA 02139

[2]Department of Chemistry, Massachusetts Institute of Technology, Cambridge, MA 02139

*email: hjkulik@mit.edu

**Contents**

**Table S1.** Element filtering for CSD ligands. Excluded elements have an abundance < 0.1% in CSD ligands of mononuclear octahedral complexes.

| allowed | H, B, C, N, O, F, Si, P, S, Cl, Br, I |
|---|---|
| excluded | Li, Be, Al, Ga, As, Se, Sn, Sb, Te, Tl |

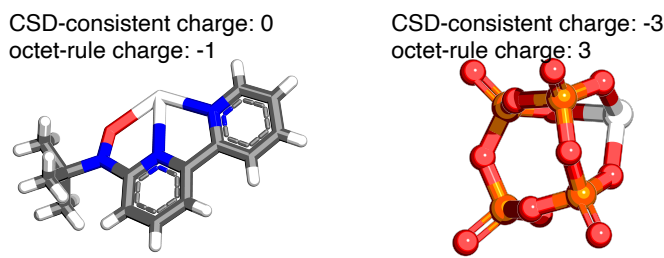

**Figure S1.** Examples where the octet rule gives incorrect ligand charge for a radical[1] (left, from a TMC with refcode WEQZOI) and a small cluster-containing ligand[2] (right, from a TMC with refcode EHUCUG). The CSD-consistent charge was verified by reviewing the original publication associated with the structure. Atoms are colored as the following, gray for C, blue for N, red for O, orange for P, white for H. For each ligand, a placeholder for metal to which the coordinating atoms connect is also shown as white.

**Table S2.** MR diagnostics calculation attrition counts and reasons.

| type | count | reason |
|---|---|---|
| Zero-temperature DFT (B3LYP, BLYP, B1LYP, PBE, and PBE0) | 521 | SCF convergence issue |
| Finite-temperature DFT (B3LYP and PBE) | 1 | SCF convergence issue |
| CASSCF | 29 | CASSCF convergence issue or exceeding the time limit of 48 hours. |
| CCSD(T) | 36 | Matrix-driven configuration interaction (MDCI) convergence issue or exceeding the time limit of 48 hours. |

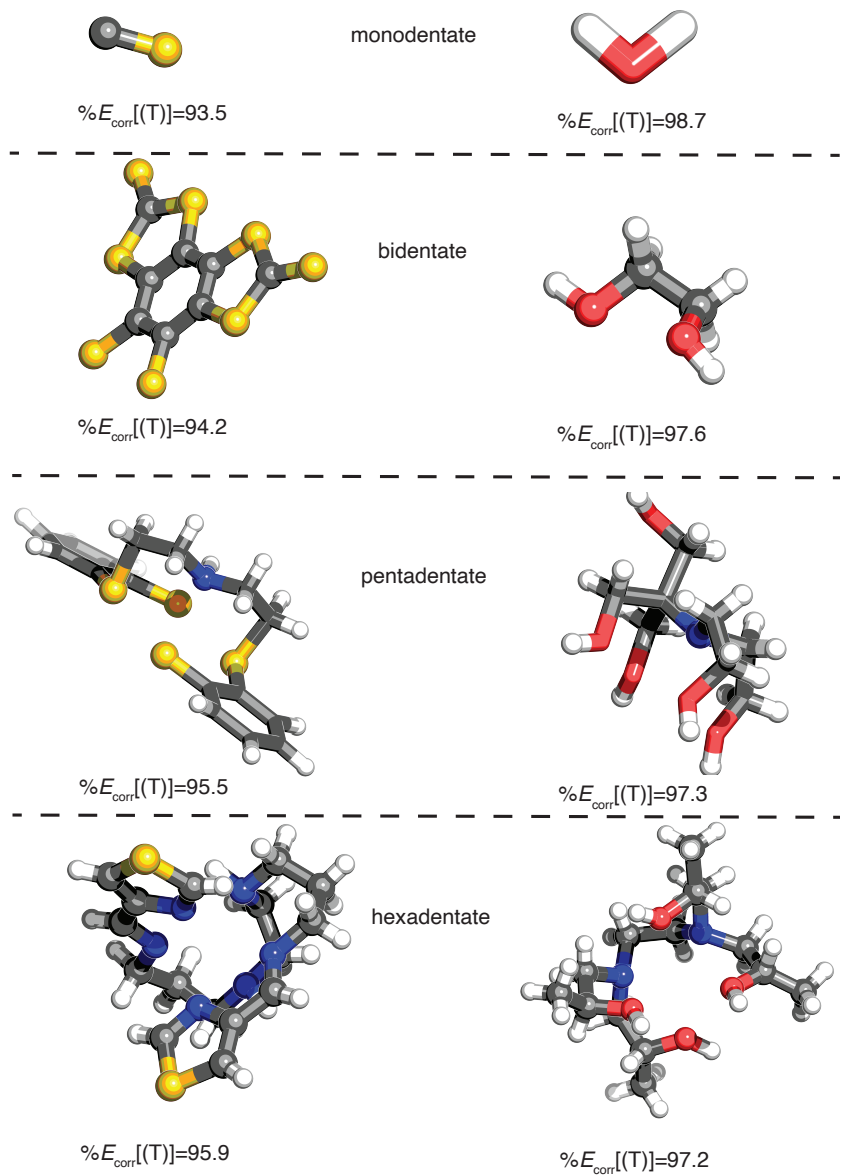

**Figure S2.** Ligands that have the strongest (left) or weakest (right) MR character for denticity 1, 2, 5, or 6 ligands (as labeled inset) from the *OctLig* set that were chosen as fragments to assemble complexes in *extTMC* set.

**Table S3.** Ligand symmetries considered in the *extTMC* set. For each symmetry, there are four choices for the metal core (i.e., singlet Fe(II), quintet Fe(II), doublet Fe(III), and sextet Fe(III)), resulting in 43*4=172 complexes in total in the *extTMC* set. We used a short-hand notation, MR$^{denticity}$ for the ligand. For example, a strong MR character bidentate ligand is S$^2$. Thus, a ligand symmetry made of two bidentate ligands of strong MR character and two monodentate ligands of weak MR character is (S$^2$)$_2$(W$^1$)$_2$. We use the bracket (i.e., []) to group isomers together. For symmetries made of two bidentate and two monodentate ligands, we require the two bidentate ligands reside in the equatorial plane.

| description | count | symmetries |
|---|---|---|
| Six monodentate | 10 | (S$^1$)$_6$, (S$^1$)$_5$(W$^1$)$_1$, [*cis* (S$^1$)$_4$(W$^1$)$_2$, *trans* (S$^1$)$_4$(W$^1$)$_2$], [*fac* (S$^1$)$_3$(W$^1$)$_3$, *mer* (S$^1$)$_3$(W$^1$)$_3$], [*cis* (S$^1$)$_2$(W$^1$)$_4$, *trans* (S$^1$)$_2$(W$^1$)$_4$], (S$^1$)$_1$(W$^1$)$_5$, (W$^1$)$_6$ |
| One bidentate + four monodentate | 18 | (S$^2$)$_1$(S$^1$)$_4$, [(S$^2$)$_1$(S$^1$)$_3$(W$^1$)$_1$, (S$^2$)$_1$(W$^1$)$_1$(S$^1$)$_3$], [(S$^2$)$_1$(W$^1$)$_2$(S$^1$)$_2$, (S$^2$)$_1$(W$^1$)$_1$(S$^1$)$_1$(W$^1$)$_1$(S$^1$)$_1$, (S$^2$)$_1$(S$^1$)$_2$(W$^1$)$_2$], [(S$^2$)$_1$(S$^1$)$_1$(W$^1$)$_3$, (S$^2$)$_1$(W$^1$)$_3$(S$^1$)$_1$], (S$^2$)$_1$(W$^1$)$_4$, (W$^2$)$_1$(S$^1$)$_4$, [(W$^2$)$_1$(S$^1$)$_3$(W$^1$)$_1$, (W$^2$)$_1$(W$^1$)$_1$(S$^1$)$_3$], [(W$^2$)$_1$(W$^1$)$_2$(S$^1$)$_2$, (W$^2$)$_1$(W$^1$)$_1$(S$^1$)$_1$(W$^1$)$_1$(S$^1$)$_1$, (W$^2$)$_1$(S$^1$)$_2$(W$^1$)$_2$], [(W$^2$)$_1$(S$^1$)$_1$(W$^1$)$_3$, (W$^2$)$_1$(W$^1$)$_3$(S$^1$)$_1$], (W$^2$)$_1$(W$^1$)$_4$, |
| Two bidentate + two monodentate | 9 | (S$^2$)$_2$(S$^1$)$_2$, (S$^2$)$_2$(W$^1$)$_2$, (S$^2$)$_2$(S$^1$)$_1$(W$^1$)$_1$ (W$^2$)$_2$(S$^1$)$_2$, (W$^2$)$_2$(W$^1$)$_2$, (W$^2$)$_2$(S$^1$)$_1$(W$^1$)$_1$, (S$^2$)$_1$(W$^2$)$_1$(S$^1$)$_2$, (S$^2$)$_1$(W$^2$) (W$^1$)$_2$, (S$^2$)$_1$(W$^2$) (S$^1$)$_1$(W$^1$)$_1$ |
| One pentadentate + one monodentate | 4 | (S$^5$)$_1$(S$^1$)$_1$, (S$^5$)$_1$(W$^1$)$_1$, (W$^5$)$_1$(S$^1$)$_1$, (W$^5$)$_1$(W$^1$)$_1$ |
| One hexadentate | 2 | (S$^6$)$_1$, (W$^6$)$_1$ |

**Table S4.** Metals (M), oxidation states (ox), and spin states considered in this work. Cases where the spin state is not included are shown with "--".

| $d$ electron configuration | M(ox) | High-spin state | Intermediate-spin state | Low-spin state |
|---|---|---|---|---|
| $d^3$ | Cr(III) | -- | quartet | doublet |
| $d^4$ | Mn(III)/Cr(II) | quintet | -- | singlet |
| $d^5$ | Fe(III)/Mn(II) | sextet | -- | doublet |
| $d^6$ | Co(III)/Fe(II) | quintet | -- | singlet |
| $d^7$ | Co(II) | -- | quartet | doublet |

**Table S5.** Geometry cutoffs used in this work, which are the same as our prior work[3]. These geometry checks include: coordination number must be 6, mean and max. deviation of connecting-atom–metal–connecting-atom angles, max. deviation of any metal–ligand bond length, max. deviation of metal–ligand bond lengths for equatorial ligands, max. root mean square deviation (RMSD) of the ligand from its initial structure, and mean and max. deviation of an expected linear ligand from a 180° angle (which are only computed for linear ligands).

| coordination number | | | |
|---|---|---|---|
| 6 | | | |
| **first coordination sphere shape** | | | |
| mean($\Delta\theta(C_i\text{-}M\text{-}C_j)$) | max($\Delta\theta(C_i\text{-}M\text{-}C_j)$) | max($\Delta d$) | max($\Delta d_{eq}$) |
| 12° | 22.5° | 1.0 Å | 0.25 Å |
| **ligand distortion metrics** | | | |
| max(RMSD) | | mean($\Delta\theta(M\text{-}A\text{-}B)$) | max($\Delta\theta(M\text{-}A\text{-}B)$) |
| 0.30 Å | | 20° | 28° |

**Table S6.** Summary of MR diagnostics grouped by type and method used. The only change from prior work[4] is that $D_1$ and $D_2$ diagnostics were removed considering their high linear correlations with some of the diagnostics ($D_1$ with $T_1$ and $D_2$ with $C_0^2$) and max($t_1$) was added as in our recent work[31].

| Diagnostic | Method | Type | Extended description |
| --- | --- | --- | --- |
| $B_1$[5] | DFT | TAE | Differences in total atomization energy for BLYP and B1LYP (25% exchange) divided by number of pairs of bonded atoms |
| $A_{25}$[PBE][6] | DFT | TAE | 4x the difference in TAE[PBE] and TAE[PBE0] (25% exchange) divided by TAE[PBE] |
| $I_{ND}$[PBE][7] | DFT | occupations | Estimation of non-dynamical contribution from finite-temperature DFT with PBE functional (T = 5000 K) |
| $r_{ND}$[PBE][8] | DFT | occupations | ratio of FT-DFT $I_{ND}$ from PBE to the sum of $I_{ND}$ with the dynamical term, $I_D$ |
| $I_{ND}$[B3LYP][7] | DFT | occupations | Estimation of non-dynamical contribution from finite-temperature DFT with B3LYP functional (T = 9000 K) |
| $r_{ND}$[B3LYP][8] | DFT | occupations | ratio of FT-DFT $I_{ND}$ from B3LYP to the sum of $I_{ND}$ with the dynamical term, $I_D$ |
| $n_{HOMO}$[MP2][6, 9] | MP2 | occupations | MP2 highest occupied natural orbital occupation |
| $n_{LUMO}$[MP2][6, 9] | MP2 | occupations | MP2 lowest unoccupied natural orbital occupation |
| $T_1$[10] | CCSD | excitations | Frobenius norm of the single-excitation amplitude vector normalized by the square root of the number of electrons in CCSD |
| max($t_1$)[11] | CCSD | excitations | The largest eigenvalue of the matrix derived from the single-excitation amplitudes. |
| %TAE[(T)][12] | CCSD(T) | TAE | Percent difference in TAE from CCSD vs. CCSD(T) |
| $C_0^2$[14][10, 13] | CASSCF | occupations | CASSCF leading coefficient CSF at an active space of 14 orbitals |
| $n_{HOMO}$[14][6, 14] | CASSCF | occupations | CASSCF highest occupied natural orbital occupation at an active space of 14 orbitals |
| $n_{LUMO}$[14][6, 14] | CASSCF | occupations | CASSCF lowest unoccupied natural orbital occupation at an active space of 14 orbitals |

**Table S7.** Default convergence parameters used in self-consistent field calculations (i.e., HF, DFT, and CCSD(T)) for ORCA 4.2.1[15].

| Software | Energy convergence threshold (Ha) | DIIS error (Ha) | Maxiter |
| --- | --- | --- | --- |
| ORCA | 1e-6 | 1e-6 | 125 |

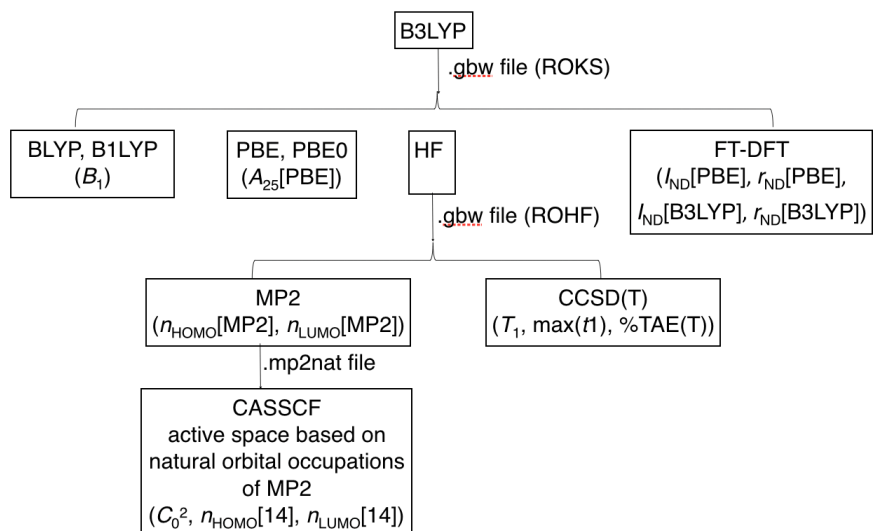

**Figure S3.** Workflow for computing 14 MR diagnostics following our previous established procedure[16-18]. All calculations were performed with ORCA 4.2.1[15].

**Text S1.** Workflow for the calculations of 14 MR diagnostics.
We converged a B3LYP calculation and used it to initialize both DFT calculations with other density functional approximations (i.e., BLYP, B1LYP, PBE, and PBE0) as well as HF calculations. This ensured we converged a consistent electronic state across multiple calculations and simultaneously saved computational time. The converged HF wavefunction was then used for MP2 and CCSD(T) calculations following an established procedure in our previous work[17] (Figure S3). All MP2 natural orbitals with occupations between 0.02 and 1.98 were initially selected for the CASSCF active space. If these thresholds led to an active space that would produce greater than $1 \times 10^7$ configuration state functions, the thresholds for occupation-based inclusion were tightened, limiting the maximum number of orbitals in the active space to 19 across the *OctLig* set. During the computation of total atomization energy (TAE)-based diagnostics, we assumed homolytic dissociation for the atoms in the ligands. We chose the percentage of correlation energy recovered by CCSD compared to CCSD(T) (i.e., %$E_{corr}$[(T)]) as a figure of merit, since we observed good correspondence of %$E_{corr}$[(T)] and %$E_{corr}$[T] in both organic molecules[4] and TMCs[16] from our previous work. For the *extTMC* and *divTMC* sets, we used DLPNO-CCSD(T) as a proxy of the canonical CCSD(T) to compute %$E_{corr}$[(T)] due to the sufficient accuracy of DLPNO-CCSD(T) and demanding computational cost of canonical CCSD(T)[19] (Table S8).

**Table S8.** Default parameters used for DLPNO-CCSD(T) calculations when "TightPNO" is specified for ORCA 4.2.1[15], which has been shown to approach to canonical CCSD(T) accuracy[19].

| $T_{CutPairs}$ | $T_{CutPNO}$ | $T_{CutMKN}$ | MP2 pair treatment |
|---|---|---|---|
| 1e-5 | 1e-7 | 1e-4 | fully iterative |

**Table S9**. Range of hyperparameters sampled for ANN models trained from scratch with Hyperopt[20]. The lists in the architecture row can refer to two or three hidden layers (i.e., the number of items in the list), and the number of nodes in each layer are denoted as elements of the list. The built-in Tree of Parzen Estimator algorithm in Hyperopt was used for the hyperparameter selection process.

| Architecture | ([128], [256], [512], [128, 128], [256, 256], [512, 512], [128, 128, 128], [256, 256, 256], [512, 512, 512]) |
|---|---|
| L2 regularization | [1e-6, 1] |
| Dropout rate | [0, 0.5] |
| Learning rate | [1e-5, 1e-3] |
| Beta1 | [0.75, 0.99] |
| Batch size | [16, 32, 64, 128, 256] |

**Table S10.** Number of training and test data points in each partitioning scheme and their corresponding percentage in parenthesis.

| scheme | train | test |
|---|---|---|
| random | 684 (80%) | 171 (20%) |
| excluding 50 ligands | 625 (73%) | 230 (27%) |
| 2p/3p | 703 (82%) | 152 (18%) |

**Text S2.** Pseudocode for iterative CSD-consistent charge assignment.

```
charge_dict, added_new_ligands = dict(), True
unassigned_complexes, assigned_complexes = all_complexes, set()
while added_new_ligands:
      added_new_ligands = False
      loop through unassigned_complexes:
            if only one ligand in the complex is absent in charge_dict:
                  derive the charge of this ligand and update charge_dict
                  move this complex from unassigned to assigned
                  added_new_ligands = True
      check consistency in charge_dict to rule out failed cases
```

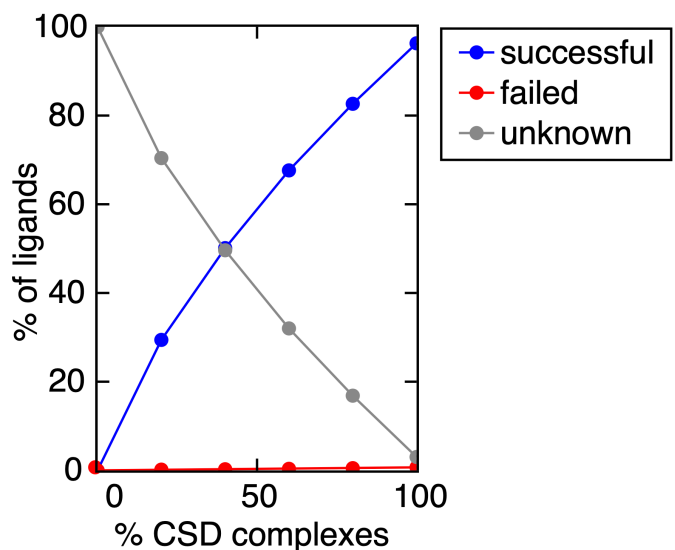

**Figure S4.** Percentage of ligands vs. percentage of CSD complexes included in the task of iterative CSD-consistent charge assignment. The results of ligand charge assignment are grouped as successful (blue), failed (red), or unknown (gray). The x-axis represents the extent of a randomly sampled subset of the entire set (i.e., 100%) of mononuclear octahedral transition metal complexes in CSD.

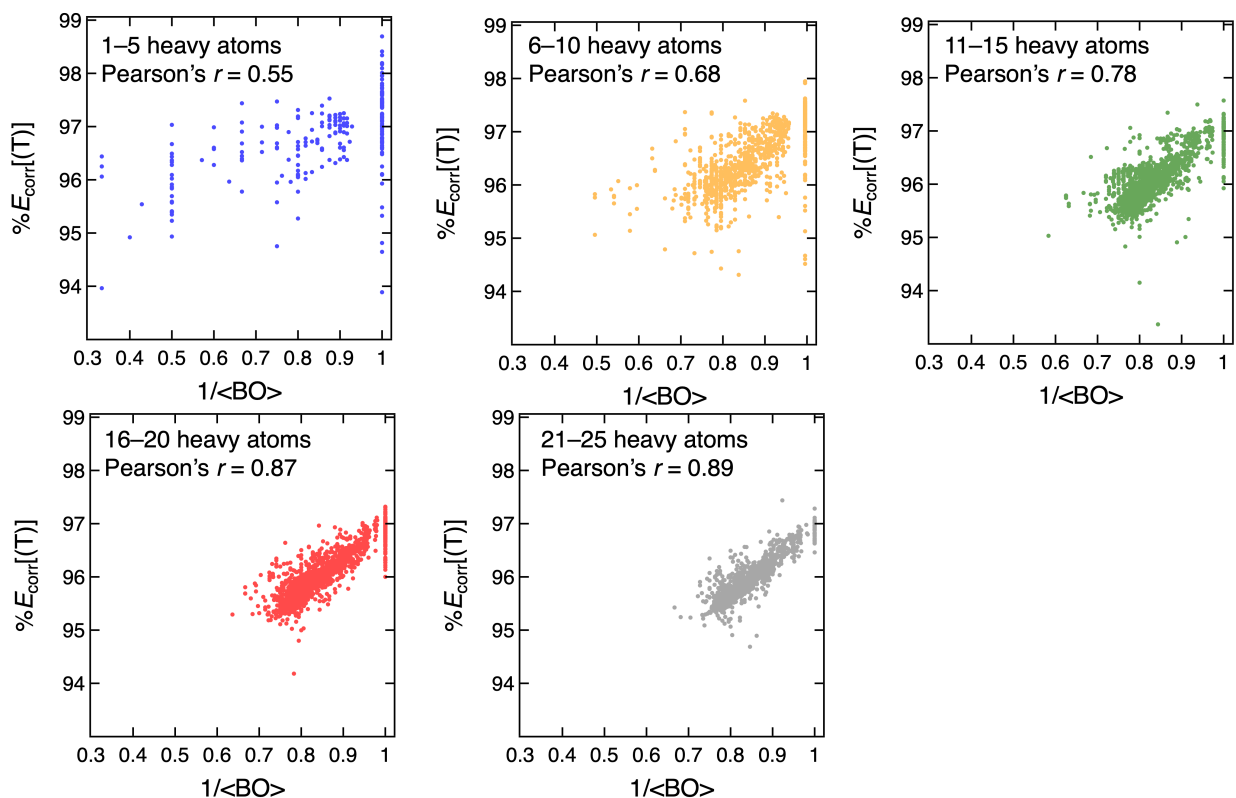

**Figure S5.** $\%E_{corr}[(T)]$ versus the inverse average bond order of molecules of different sizes.

**Table S11.** Detailed information for select ligands with triple bonds

| formula | SMILES | %$E_{corr}$[(T)] |
|---|---|---|
| CS | [C-]#[S+] | 93.6 |
| CN$^-$ | [C-]#N | 95.0 |
| C$_4$H$^-$ | C#CC#[C-] | 95.2 |

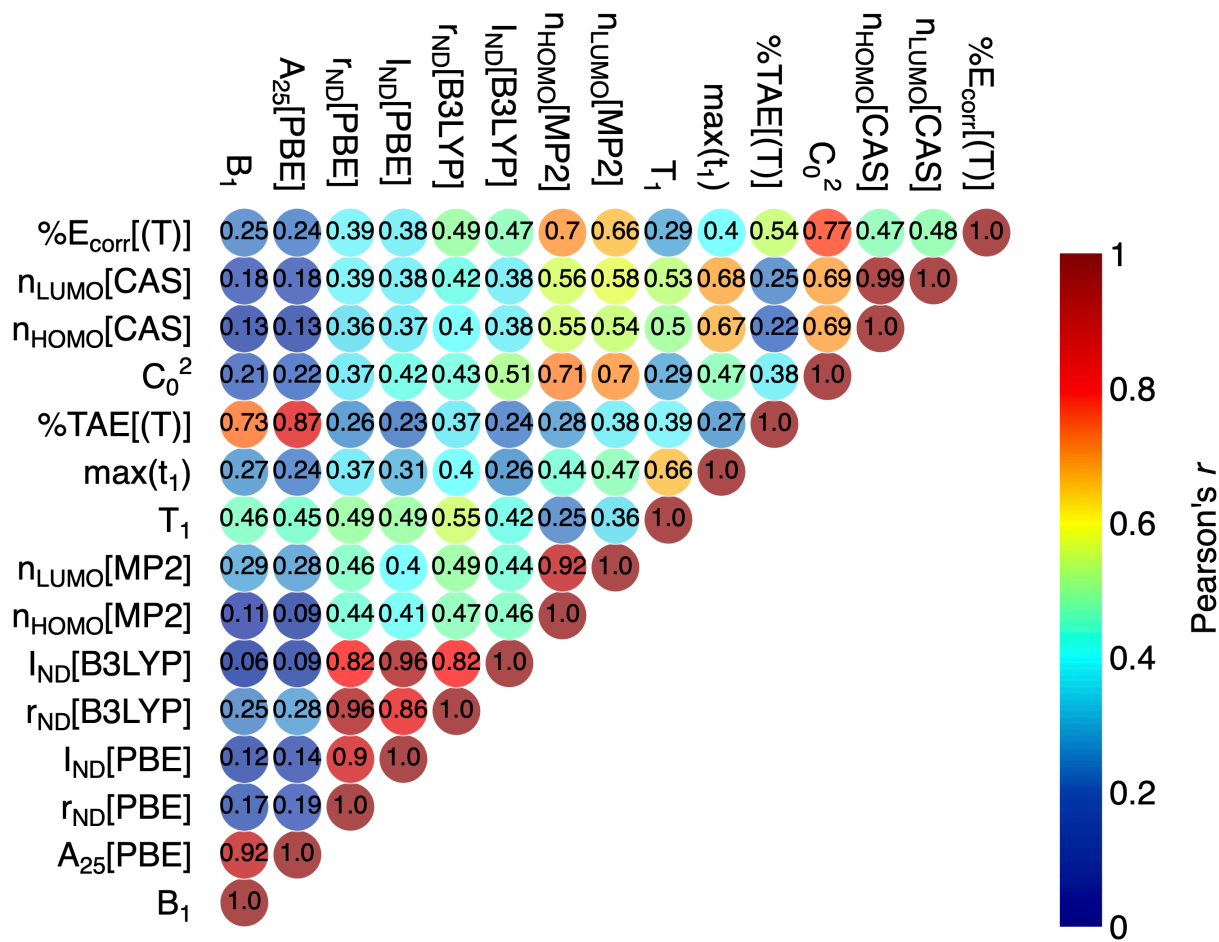

**Figure S6.** An upper triangular matrix of unsigned Pearson's $r$ for pairs of MR diagnostics and %$E_{corr}$[(T)] on the set of *OctLig*. For each pair, the circle is colored by the unsigned Pearson's $r$ and the $r$ value is explicitly shown.

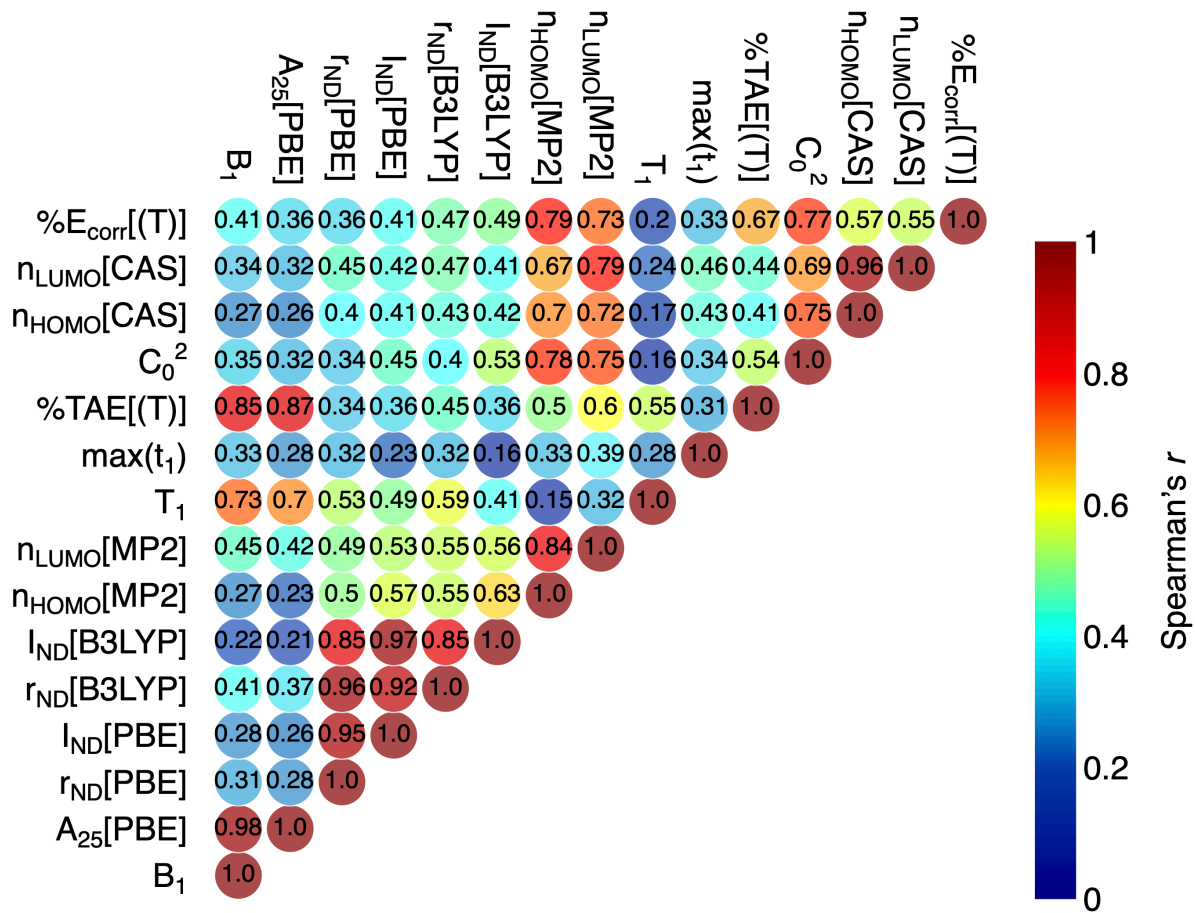

**Figure S7.** An upper triangular matrix of Spearman's $r$ for pairs of MR diagnostics and %$E_{corr}$[(T)] on the set of *OctLig*. For each pair, the circle is colored by the Spearman's $r$ and the $r$ value is explicitly shown.

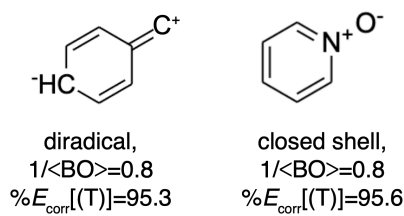

diradical,
1/⟨BO⟩=0.8
%$E_{corr}$[(T)]=95.3

closed shell,
1/⟨BO⟩=0.8
%$E_{corr}$[(T)]=95.6

**Figure S8.** 2D chemical structure and detailed information of phenylmethylidyne ($C_7H_5$, left) and pyridine-N-oxide ($C_5H_5NO$, right).

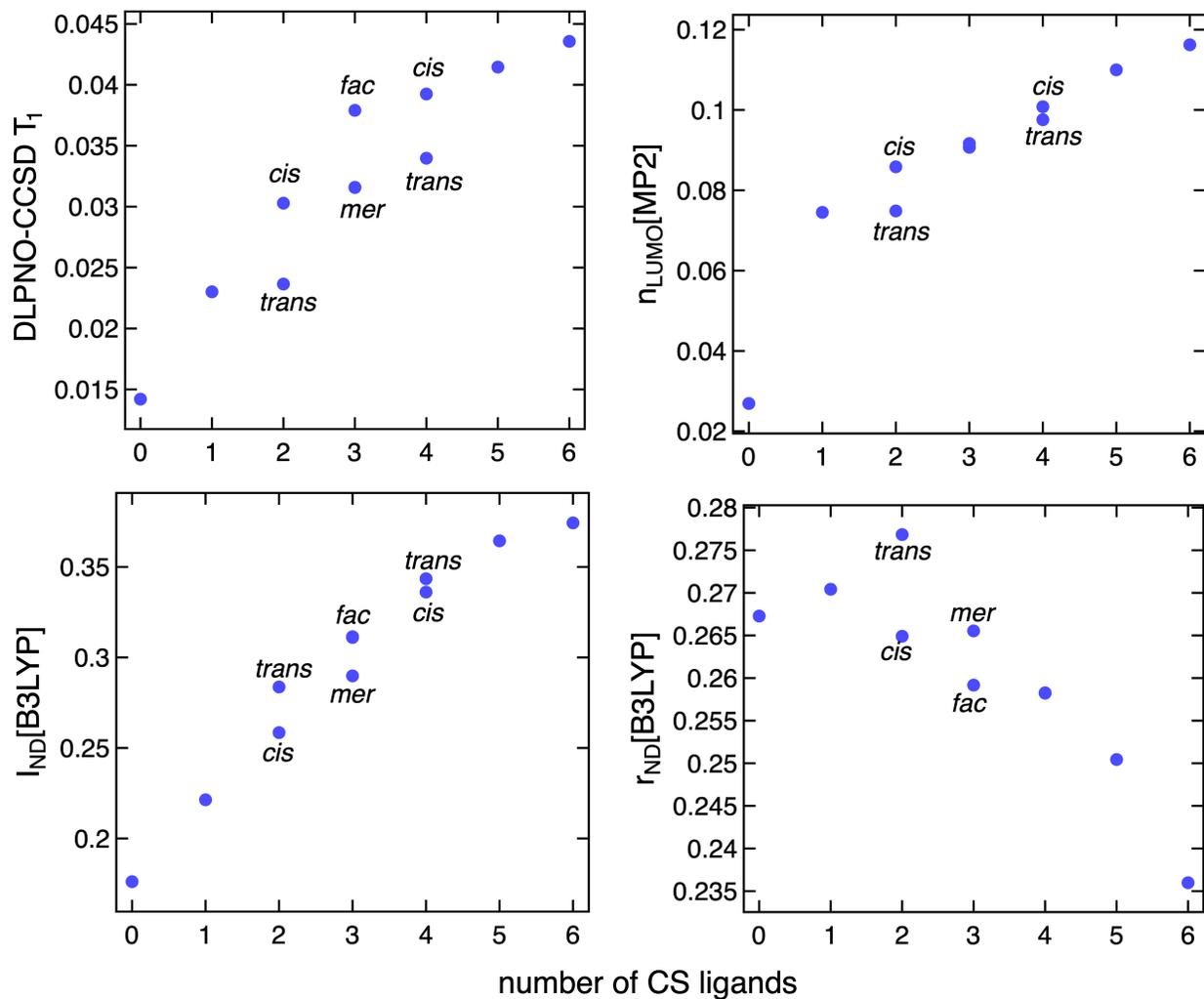

**Figure S9.** MR diagnostics versus the number of CS ligands in LS Fe(II)(CS)$_n$(H$_2$O)$_{6-n}$. The type of isomer is annotated for n=2, 3, and 4. All other diagnostics that are not shown are similarly monotonically increasing with CS ligands. See Supporting Information .zip file.

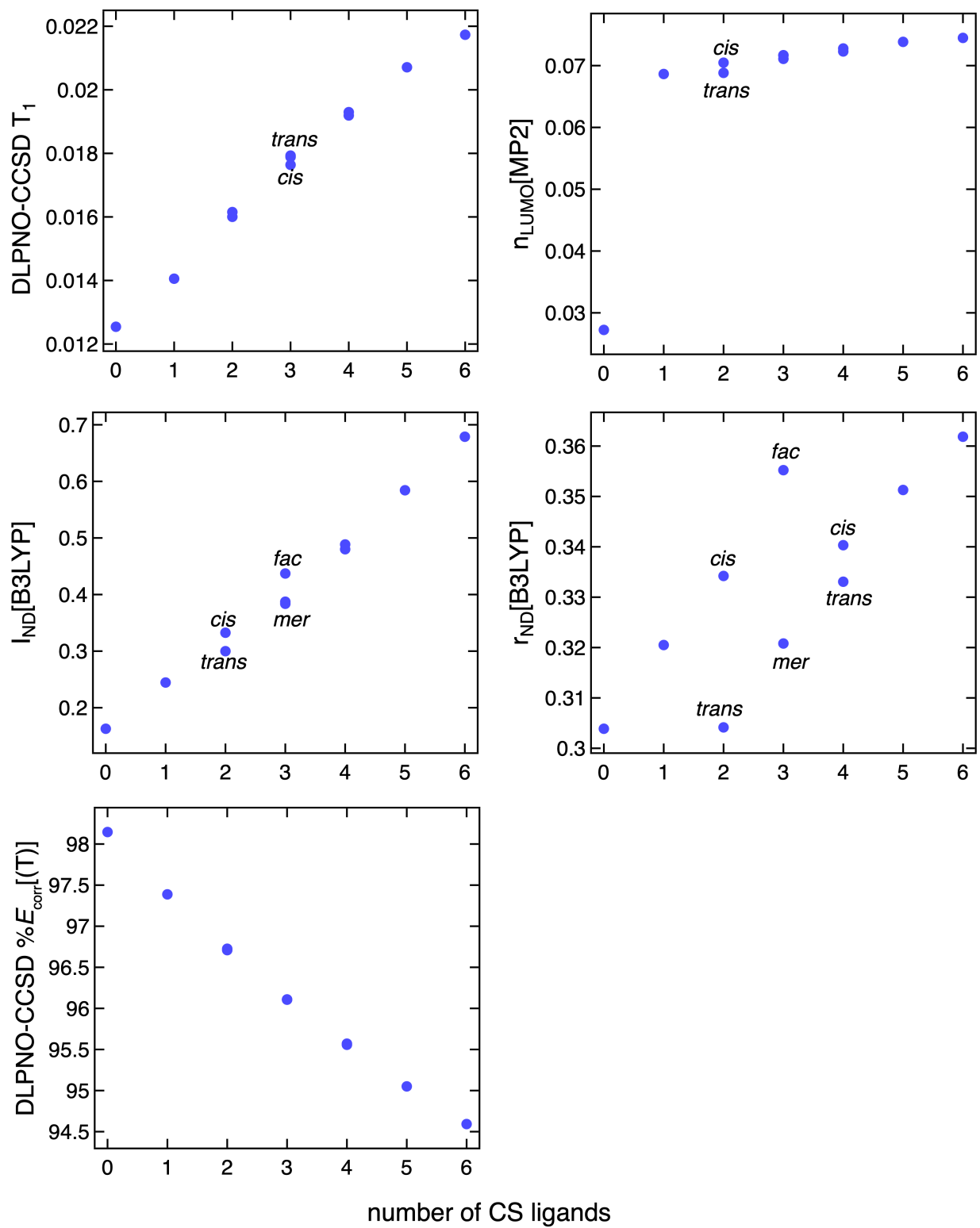

**Figure S10.** MR diagnostics versus the number of CS ligands in HS Fe(II)(CS)$_n$(H$_2$O)$_{6-n}$. The type of isomer is annotated for n=2, 3, and 4. All other diagnostics that are not shown are similarly monotonically increasing with CS ligands. See Supporting Information .zip file.

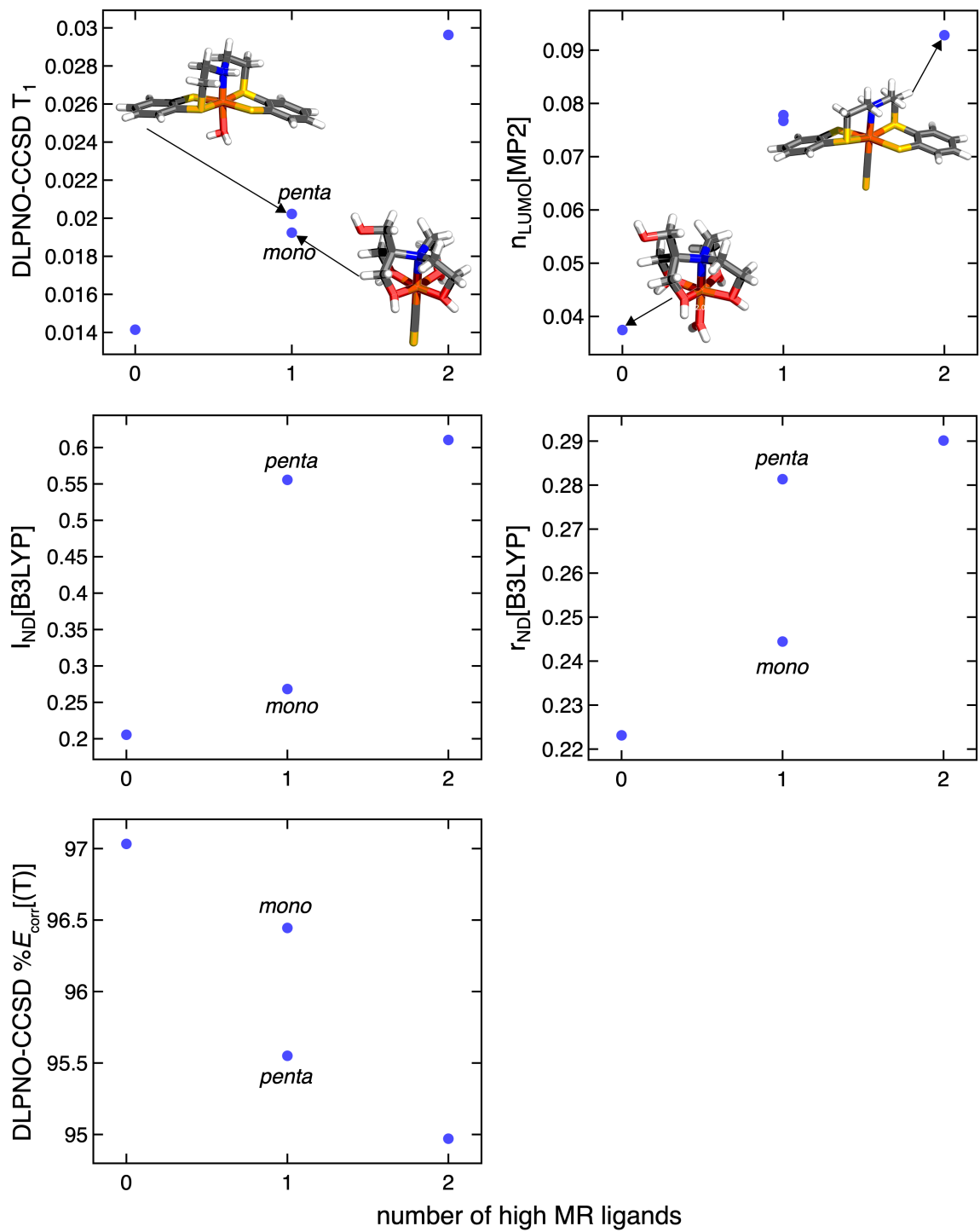

**Figure S11.** MR diagnostics versus the number of high MR ligands in LS Fe(II)($C_{14}H_{17}NS_4$)$_m$($C_8H_{19}NO_5$)$_{1-m}$(CS)$_n$($H_2O$)$_{1-n}$ (m=0 or 1; n=0 or 1). The number of high MR ligands is m+n. Each complex is shown as insets. "mono" corresponds to the complex where the high MR character ligand is monodentate (i.e., m=0 and n=1), and "penta" corresponds to the complex where the high MR character ligand is pentadentate (i.e., m=1 and n=0). Atoms are colored as follows: orange for Fe, gray for C, yellow for S, red for O, blue for N, and white for H.

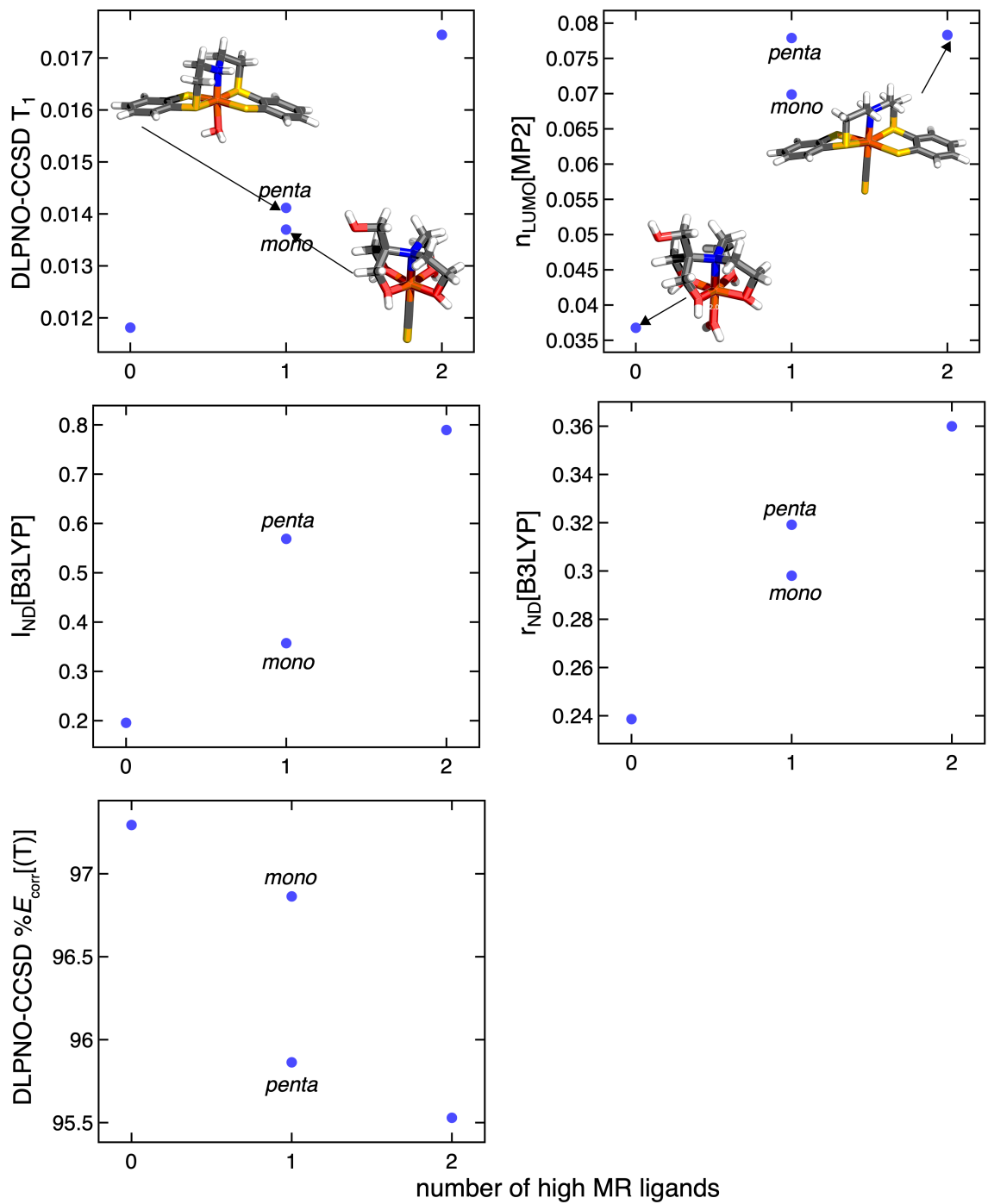

**Figure S12.** MR diagnostics versus the number of high-MR ligands in HS $Fe(II)(C_{14}H_{17}NS_4)_m(C_8H_{19}NO_5)_{1-m}(CS)_n(H_2O)_{1-n}$ (m=0 or 1; n=0 or 1). The number of high-MR ligands is m+n. Each complex is shown as insets. "mono" corresponds to the complex where the high MR character ligand is monodentate (i.e., m=0 and n=1), and "penta" corresponds to the complex where the high MR character ligand is pentadentate (i.e., m=1 and n=0). Atoms are colored as follows: orange for Fe, gray for C, yellow for S, red for O, blue for N, and white for

Page S17

H. All other diagnostics that are not shown are similarly monotonically increasing with number of high MR ligands. See Supporting Information .zip file.

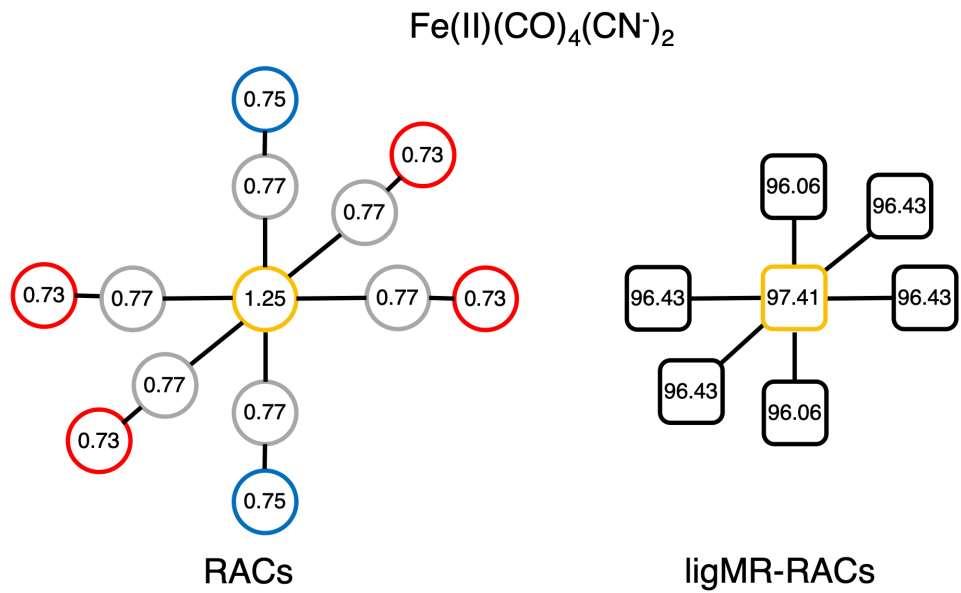

**Figure S13.** 2D molecular graph for LS Fe(II)(CO)$_4$(CN$^-$)$_2$ used for generating RACs (left) and ligMR-RACs (right). The covalent radii are annotated as the representative atomic property (left) and %$E_{corr}$[(T)] is used as the representative MR diagnostic of ligands for ligMR-RACs (right). Circles are used to represent atoms and squares for fragments (i.e., a metal center or ligand) in TMCs. Atoms are colored as follows: orange for Fe, gray for C, red for O, and blue for N.

**Text S3.** Extended description of RACs and ligMR-RACs
We have introduced a systematic approach to featurize molecular inorganic complexes that blends metal-centric and whole-complex topological properties in a feature set referred to as revised autocorrelation functions (RACs).[21] These RACs, variants of graph autocorrelations (ACs),[22] are sums of products and differences of atomic properties, i.e., electronegativity ($\chi$), nuclear charge ($Z$), topology ($T$), covalent radius ($S$), and identity ($I$). Recently, we incorporated the group number ($G$) as an additional atomic property for better generalization across rows in the periodic table. We name these new sets of descriptors eRACs[23]. Standard ACs are defined as

$$P_d = \sum_i \sum_j P_i P_j \delta(d_{ij}, d)$$

where $P_d$ is the AC for property $P$ at depth $d$, $\delta$ is the Dirac delta function, and $d_{ij}$ is the bond-wise path distance between atoms $i$ and $j$.

In our approach, we have five types of RACs:
- $_{all}^{f}P_d$: standard ACs start on the full molecule (*f*) and have all atoms in the scope (all).
- $_{ax}^{f}P_d$ and $_{eq}^{f}P_d$: restricted-*scope* ACs that start on the full molecule (*f*) and separately evaluate axial or equatorial ligand properties

$$_{ax/eq}^{\ \ \ f}P_d = \frac{1}{|ax/eq\ ligands|} \sum_i^{n_{ax/eq}} \sum_i^{n_{ax/eq}} P_i P_j \delta(d_{ij}, d)$$

where $n_{ax/eq}$ is the number of atoms in the corresponding axial or equatorial ligand and properties are averaged within the ligand subtype.

- $_{all}^{mc}P_d$: restricted-scope, metal-centered (mc) descriptors that start on the metal center (mc) and have all atoms in the scope (all), in which one of the atoms, $i$, in the $i,j$ pair is a metal center:

$$_{all}^{mc}P_d = \sum_i^{mc} \sum_i^{all} P_i P_j \delta(d_{ij}, d)$$

- $_{ax}^{lc}P_d$ :and $_{ax}^{lc}P_d$: restricted-scope, metal-proximal ACs that start on a ligand-centered (lc) and separately evaluate axial or equatorial ligand properties, in which one of the atoms, $i$, in the $i,j$ pair is the metal-coordinating atom of the ligand:

$$_{ax/eq}^{\ \ \ lc}P_d = \frac{1}{|ax/eq\ ligands|} \frac{1}{|lc|} \sum_i^{lc} \sum_i^{n_{ax/eq}} P_i P_j \delta(d_{ij}, d)$$

We also modify the AC definition, $P'$, to property differences rather than products for a minimum depth, $d$, of 1 (as the depth-0 differences are always zero):

$$_{ax/eq/all}^{\ \ \ lc/mc}P'_d = \sum_i \sum_i^{lc\ or\ mc\ scope} P_i P_j \delta(d_{ij}, d)$$

where scope can be axial, equatorial, or all ligands.

Here, we adapt RACs to encode the MR character of a TMC through the MR diagnostics of its fragments (i.e., metal center and ligands). Since most of the MR diagnostics cannot be localized to individual atoms, we first convert the 2D molecular graph to a fragment-based graph, where the MR diagnostics of each fragment have been pre-computed (Figure S13). To compute ligMR-RACs for a TMC, we follow the same routine in RACs, but simply operating on this fragment-based graph and replacing the atomic property in RACs by ligand MR diagnostics. Since the maximum bond depth ($d$) is 2 in fragment-based graph for all TMCs, we truncate ligMR-RACs at depth 2 as well. We remove four MR diagnostics (i.e., $I_{ND}$[B3LYP], $r_{ND}$[B3LYP], $n_{HOMO}$[MP2], and $n_{HOMO}$[14]) that have Pearson's r coefficients larger than 0.95 with at least one of the remaining diagnostics, leaving 10 diagnostics and $\%E_{corr}$[(T)] (Figure S6). Due to the constraints on our fragment-based molecular graph, $_{all}^{f}P_d$ can have $d$ from 0 to 2, $_{all}^{mc}P_d$ with $d$ as 0 or 1, $_{eq/ax}^{\ \ \ f}P_d$ and $_{eq/ax}^{\ \ \ lc}P_d$ with $d$ as 0, and $_{all}^{mc}P'_d$ with $d$ as 1. This leads to 11*10 = 110 features in total. However, three TAE-based diagnostics (i.e. $B_1$, $A_{25}$[PBE], and %TAE[(T)]) for the metal are always zero, resulting in an additional 3*3=9 features that are zero. Therefore, we have 101 non-zero ligMR-RACs for a TMC.

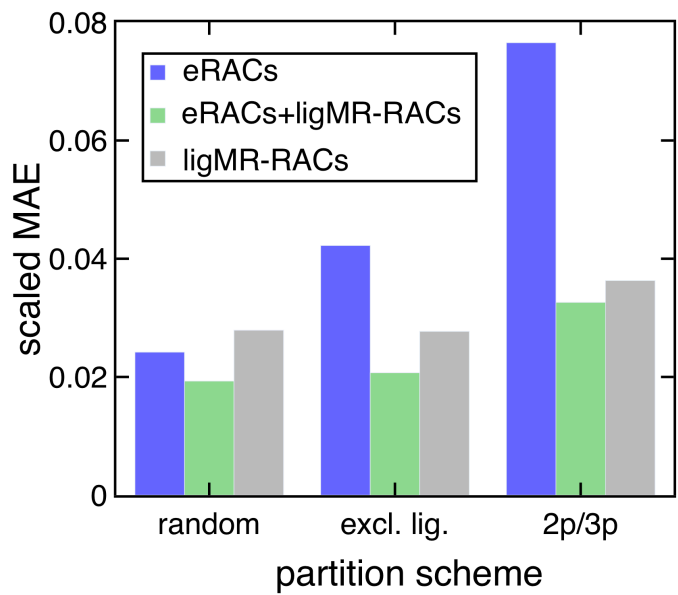

**Figure S14.** Scaled MAE for ML predicted DLPNO-CCSD %$E_{\text{corr}}$[(T)] with three partitioning schemes of using only eRACs (blue), both eRACs and ligMR-RACs (green), and only ligMR-RACs (gray). Results are obtained on the set-aside test set of *divTMC* set.

**Table S12.** ANN model performance with different feature sets and partitioning schemes.

|  | random | | | exclude ligands | | | 2p/3p | | |
| --- | --- | --- | --- | --- | --- | --- | --- | --- | --- |
|  | eRACs | eRACs + ligMR-RACs | ligMR-RACs | eRACs | eRACs + ligMR-RACs | ligMR-RACs | eRACs | eRACs + ligMR-RACs | ligMR-RACs |
| **$R^2$** | 0.95 | 0.96 | 0.91 | 0.81 | 0.96 | 0.93 | 0.51 | 0.93 | 0.89 |
| **scaled MAE** | 0.024 | 0.019 | 0.028 | 0.042 | 0.021 | 0.028 | 0.077 | 0.032 | 0.036 |